\documentclass{svproc}

\usepackage{cite}
\usepackage{url}  
\PassOptionsToPackage{hyphens}{url}
\usepackage[hidelinks]{hyperref}
\hypersetup{breaklinks=true}
\usepackage{multicol}

\usepackage{enumitem}
\usepackage{lipsum}
\usepackage[dvipsnames]{xcolor}

\newcommand{\refsec}[1]{Section~\ref{#1}}
\newcommand{\reffig}[1]{Figure~\ref{#1}}
\newcommand{\reftbl}[1]{Table~\ref{#1}}

\newcommand{\etal}{~et~al.}
\newcommand{\ttt}[1]{{\small \texttt{#1}}}

\usepackage{graphicx}
\graphicspath{{images/}}

\usepackage{caption}
\usepackage{subcaption}
\usepackage[dvipsnames]{xcolor}

\DeclareCaptionLabelSeparator{periodspace}{.\quad}
\def\BibTeX{{\rm B\kern-.05em{\sc i\kern-.025em b}\kern-.08em
    T\kern-.1667em\lower.7ex\hbox{E}\kern-.125emX}}

\begin{document}
\mainmatter              
\title{Does More Bandwidth Really Not Matter (Much)?}
\titlerunning{Does More Bandwidth Really Not Matter (Much)?}  
%
\author{Seraj Al Mahmud Mostafa\inst{1} \and Mike P. Wittie\inst{1} \and Utkarsh Goel\inst{2}}
\authorrunning{Seraj Mostafa et al.} 
%
%
\institute{Montana State University, Bozeman, MT, USA
\and
Akamai
}

\maketitle              

\pagestyle{plain}

\begin{abstract}

The prevailing wisdom is that more network bandwidth does not matter much and that website performance is primarily limited by network latency.
However, as mobile websites become more complex and mobile network performance improves, does this adage continue to hold?
To understand the effects of small changes in network bandwidth and latency on website performance, we propose a novel webpage characterization metrics -- Critical Path of Improvement~(CPI).
We compute CPI for 45 websites and analyze it against the network performance of four mobile ISPs in 57 US cities.
Our results show that 18\% of websites are primarily limited by bandwidth with others limited by bandwidth to some extent.
These results show that contrary to accepted wisdom, insufficient bandwidth is a limiting factor in some website/network combinations.
We also offer a discussion of approaches website developers and mobile network administrators can follow to understand and mitigate bandwidth limitations to website performance.

\keywords{Mobile web, measurement, web performance metrics}
\end{abstract}

\section{Introduction}


The evolution of mobile web services goes hand-in-hand with the improvements in mobile network performance.
As mobile networks offer lower latency and higher bandwidth, mobile websites may transfer more content to offer a richer experience, while maintaining good response times.
But what are there instances where website design and network performance do not match up well?


The prevailing wisdom is that network latency is the limiting factor to website performance.
Two key articles driving this idea came from Mike Belshe~\cite{Belshe2010More} and Ilya Grigorik~\cite{Grigorik2012Latency}.
They observe that beyond some point, adding more bandwidth results in diminishing marginal returns of website load time.
Incremental reduction of latency, on the other hand, results in corresponding improvement of website performance.
It would seem then that beyond a certain minimal bandwidth website performance is limited by network latency.

While that may have been true a decade ago for landline internet and desktop websites, much of today's web traffic involves mobile networks and smartphones.
The mobile networks themselves have undergone significant improvement in the last several years.
OpenSignal reports that from February,~2016 to January,~2020 4G latency (averaged across providers) improved from 75.5\,ms to 52.9\,ms (by 29.9\%) and bandwidth from 9.7\,Mbps to 25.6\,Mbps (by 163\%)~\cite{2020OpenSignal, 2016OpenSignal}.
At the same time, there remain pronounced regional differences in mobile network performance.
Specifically, in January, 2020 the variation of mean latency between US cities was as large as 37.8\,ms and the variation of mean bandwidth more than 28.9\,Mbps~\cite{2020OpenSignal}.
Further, as websites become more complex they may act differently in networks of varying performance.
Considering these differences, does the adage of network latency being the limiting factor website performance continue to hold?

\begin{figure}
  \centering
  \includegraphics[scale=1.2]{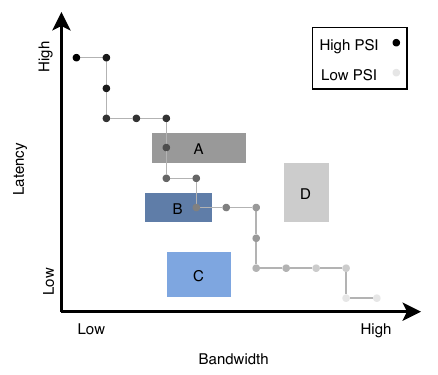}
  \caption{Illustration of CPI and network conditions.}
  \label{fig:cpi}
\end{figure}

To answer this question, we take a fresh look at characterizing mobile website network performance requirements.
We propose a new website characterization method we dub the Critical Path of Improvement~(CPI), illustrated in \reffig{fig:cpi}.
CPI allows us to understand how small changes to network bandwidth and latency affect website performance as measured by the Perceptual Speed Index~(PSI)~\cite{Gao2017Perceived}, or any other web performance metric~\cite{Furtak2019ODDness}.
We base CPI on the observation that at any given point of network performance on the bandwidth/latency plane, PSI will improve on average with small increases to bandwidth, or small reductions to latency.
\reffig{fig:cpi} illustrates PSI improvement by lighter points along the CPI.
We evaluate the impact on PSI of these small changes in network performance and record CPI as the path of the fastest PSI improvement (details in \refsec{sec:solution}).
Effectively, CPI allows us to understand when a website performance benefits more from additional bandwidth or from lower latency.

To understand whether latency or bandwidth of mobile network is the limiting factor for a given website, we analyze the relationship between a network's performance and the CPI.
In \reffig{fig:cpi}, the rectangular regions represent network performance envelopes in terms of minimum and maximum latency and bandwidth measurements.
For a network whose performance envelope lies in region D, the path to better PSI points downwards to the CPI and so via lower latency.
On the other hand, if the network performance envelope corresponds to region C, the path to the CPI and better PSI lies towards higher bandwidth.
We want to understand how common are scenarios where a website could benefit from higher bandwidth, or lower latency, than what is available in a given network deployment.
In other words, how often is website performance limited by insufficient bandwidth versus by high latency.


This paper offers the following contributions:
\begin{itemize}
\item We propose CPI -- a new website performance characterization method.
We show how developers may use CPI to understand the impact of small changes in network performance on the performance of their websites.

\item We analyse the CPI of 45 websites from the Alexa-100 list in four mobile networks across 57 US cities to identify cases where websites are bandwidth- and latency-limited.

\item We analyse regional variation in network performance to show areas of the US where network bandwidth tends to be the limiting factor to website performance.

\item We point to factors of webpage design that contribute to websites being limited by bandwidth.

\item We offer a discussion of how developers and network operators may use these results to improve the match between website design and network performance.
\end{itemize}


The rest of this paper is organized as follows.
In \refsec{sec:related_work} we discuss related work on website performance bottlenecks.
\refsec{sec:solution} details CPI measurements and comparisons with network performance.
In \refsec{sec:evaluation} we describe our measurement methodology and discuss measurement results.
Finally, we offer a discussion of the applicability of our findings to website developers and network administrators in \refsec{sec:discussion} and conclude in \refsec{sec:conclusions}.

\section{Related Work}
\label{sec:related_work}

A number of papers quantified performance bottlenecks in mobile websites.
Nejati and Balasubramanian found that device processing speed as well as delays in transferring objects on the critical path of rendering are the main causes of lower website performance on mobile devices~\cite{Nejati2016Depth}.
While some researchers have focused on reducing the computational load of rendering~\cite{Pourghassemi2019What, Ren2018Proteus}, others have focused on eliminating loading bottlenecks.


One approach to eliminating loading bottlenecks is to reorder objects on the critical path~\cite{Liu2017SWAROVsky, Furtak2019ODDness}.
Even with object reordering, Furtak\etal\ have shown that increased bandwidth improves page Speed Index by permitting large images to load more quickly~\cite{Furtak2019ODDness}.
However, others have observed that the impact of extra bandwidth is less pronounced on Page Load Time~(PLT)~\cite{Rajiullah2019Web}.
Han\etal\ have shown that extra bandwidth improves the performance of large pages~\cite{Han2016When}.
They loaded pages using MultiPath TCP~(MPTCP) over cellular and WiFi connections in parallel and showed that the loading process can use the bandwidth of the additional link effectively, when the performance of the links is comparable in terms of latency.

These results, however, focus on the impact of network performance in general, but do not consider the correspondence of specific web pages loaded in specific mobile networks.
While researchers agree that additional network bandwidth may be helpful in some cases, its not clear to what degree such cases are prevalent. 
In this paper we aim to answer that question.

\section{Critical Path of Improvement}
\label{sec:solution}

\begin{figure}
    \centering
    \begin{subfigure}{0.5\columnwidth}
        \centering
        \includegraphics[scale=1.2]{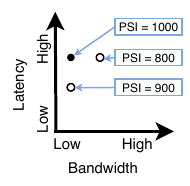}
        \caption{Evaluate candidate points.}
        \label{fig:cpi_process:1}
    \end{subfigure}%
    \begin{subfigure}{0.5\columnwidth}
        \centering
        \includegraphics[scale=1.2]{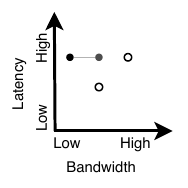}
        \caption{Add lower PSI point to CPI.}
        \label{fig:cpi_process:2}
    \end{subfigure}
    
    \caption{CPI creation process.}
    \label{fig:cpi_process}
\end{figure}

CPI measures how small changes to network performance affect website performance.
To obtain the CPI, we load a given website under different combinations of latency and download bandwidth controlled using NetEm~\cite{2011NetEm}.
For each set of network conditions we measure website PSI using PWMetrics~\cite{Irish2020PWMetrics}.

Using these tools, we measure CPI as follows:
\begin{itemize}
\item As a starting point we chose a set of atrocious network conditions, specifically bandwidth of 256\,Kbps and latency of 180\,ms.
We load a website over thus constrained link and measure its PSI.
For example, in \reffig{fig:cpi_process:1} these low bandwidth, high latency conditions might result in the PSI of 1000.

\item We then proceed to find the next CPI point.
We measure the PSI of two candidate points.
For the first candidate point we improve network latency by 10\,ms, but keep the bandwidth of the starting point.
For the second candidate point, we improve network bandwidth, but keep the latency of the starting point.
Since improvements in network bandwidth have a greater impact on website performance in the lower ranges~\cite{Belshe2010More, Grigorik2012Latency}, we double the network bandwidth up to 8\,Mbps and then increase it by a constant interval of 8\,Mbps.
We then measure the PSI of the candidate points.
In \reffig{fig:cpi_process:1} the candidate points have the PSI of 900 and 800 respectively.

\item Finally, we select the next point on the CPI as the candidate point with the lower PSI of the two, as shown in \reffig{fig:cpi_process:2}.
Its important to note that website PSI varies somewhat across loads.
When computing the CPI of the starting and candidate points we average the PSI over seven trials.
\end{itemize}

\reffig{fig:cpi} shows an example of a complete CPI measurement.
As we move along the CPI from the starting point, we can see improvements to PSI represented as lighter points.
The CPI shows how small changes to network performance improve website performance.
The CPI also shows a website's \emph{preference} for higher bandwidth, or lower latency at any point.
A CPI with more points further away from the x-axis show a preference for more bandwidth, while those with more points close to the x-axis show preference for lower latency.

We also want to consider the relationship between the CPI and the performance of a network to understand whether a website is bandwidth- or latency-limited \emph{in a given network}.
In \reffig{fig:cpi} we plot four areas: A, B, C, and D.
These areas represent some possible envelopes of a networks performance represented as minimum and maximum latency and bandwidth.
While a network's performance varies during a given time period, it will in general be bounded some maximum and minimum values, as reported by OpenSignal measurements~\mbox{\cite{2018OpenSignal, 2019OpenSignal, 2020OpenSignal}}.
As such, a network's performance constrains page PSI to a particular region of the bandwidth/latency region.

Referring to the regions in \reffig{fig:cpi} identify four cases of these constraints.
\begin{itemize}
    \item Case A: the CPI exits network performance bounds through the minimum latency boundary of area A.
    This means that page PSI would improve more if A extended to better (lower) latency, rather than towards higher bandwidth.
    \item Case B: the CPI exits network performance bounds through the maximum bandwidth boundary of area B.
    This means that page PSI would improve more if B extended to better (higher) bandwidth, rather than towards lower latency.
    \item Case C: the CPI lies to the right of the network performance bounds of area~C.
    This means that page PSI would improve more if C extended to better (higher) bandwidth, rather than towards lower latency.
    The rationale for this claim is that a CPI measurement started in area C tends towards the shown CPI and so by improving network bandwidth to the right.
    \item Case D: the CPI lies below the network performance bounds of area D.
    This means that page PSI would improve more if D extended to better (lower) latency, rather than towards higher bandwidth.
    The rationale for this claim is that a CPI measurement started in area D tends towards the shown CPI and so by improving network latency downward.
\end{itemize}

Our goal is to understand how common are the B and C cases.
Their presence indicates that web pages could benefit from additional network bandwidth, which stands in opposition to the accepted wisdom that web performance is constrained by insufficient latency.


\section{Evaluation}
\label{sec:evaluation}

Our goal is to understand the relationship between page CPI and network performance envelopes.
To frame our results, we first describe our measurement methodology.

\subsection{Measurement Methodology}

We chose 43 pages from the list of Alexa top 100 pages, excluding adult content~\cite{Alexa2020TopSites}.
To satisfy our curiosity, we also added major news sites from the authors home countries: \ttt{prothom-alo.com} and \ttt{wyborcza.pl}.
For each site, we measure the CPI using the process described in the previous section.
We load each page from within the Montana State University's wired network, while constraining the performance of the internet link with NetEm~\cite{2011NetEm}.
Using \ttt{speedtest.net} we measured the performance of our campus network to the regional ISP hosting the Content Distribution Network~(CDN) servers that serve web content.
Our network is limited to 18.14\,ms of latency and 307.64\,Mbps of bandwidth, and so NetEm constrains our network in real terms up to these numbers.


We then find the network performance regions of network providers using data from OpenSignal.
Specifically we extract 4G mobile network performance for four major network providers~(AT\&T, Sprint, T-Mobile, and Verizon) across the US for January 2020, 2019, and July 2018~\cite{2020OpenSignal, 2019OpenSignal, 2018OpenSignal} using WebPlotDigitizer~\cite{Rohatgi2019WebPlotDigitizer}.
These data contain the mean latency and bandwidth measurements in a 57 cities along with measurement confidence intervals~\cite{Boyland2019ConfidenceIntervals}.
These confidence intervals describe the likely upper and lower bounds on network performance and we treat them as the upper and lower bounds of the network performance envelopes.

\subsection{Results}

\begin{figure}
    \centering
  \includegraphics[width=0.7\columnwidth]{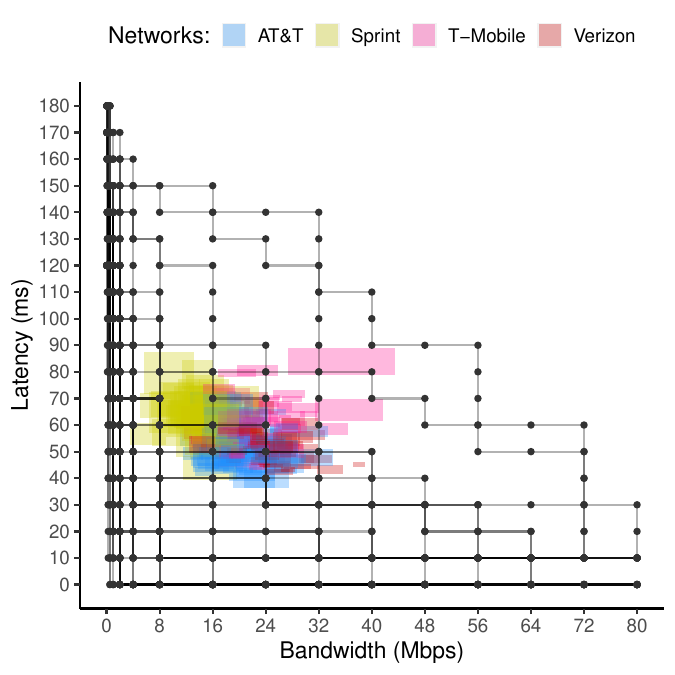}
  \caption{CPI and network conditions.}
  \label{fig:all_data}
\end{figure}

\reffig{fig:all_data} shows a summary of our dataset.
The x and y-axis respectively show the link bandwidth and latency constraints of our CPI measurements.
The figure includes CPIs for all the websites in our dataset. 
Each CPI is represented by a semi-transparent gray line.
Segments where these lines are darker represent overlaps between CPIs of different websites.
The points along CPI lines represent normalized PSI values -- we illustrate improvements in PSI more clearly further down in this section.
Finally, the figure represents network performance envelopes for the four provider networks.

Overall we observe that as the network performance improves, so does page PSI, as CPI points become lighter in the low right quadrant of \reffig{fig:all_data}.
We also see that CPIs tend to concentrate in the area 80\,ms\,-\,180\,ms and 256\,Kbps\,-\,4\,Mbps, as overlapping CPI lines become darker, but spread out at latencies lower than 60\,ms.
This tells us that while the performance of many websites initially benefits primarily from lower latency, once latency becomes sufficiently low, the benefits from improving bandwidth and latency become comparable.

We also observe that the performance of network providers tends to concentrate in a region of 40\,ms-80\,ms and \mbox{8\,Mbps-32\,Mbps}.
The CPIs of different websites cross this region, and so we expect to see that these websites are limited by latency and by bandwidth with respect to the different networks.
In the following subsections we delve deeper into this phenomenon by characterizing the relationships between individual CPIs and individual network performance regions into A, B, C, and D cases.

\subsubsection{How often are individual websites limited by bandwidth?}

\begin{figure}
    \centering
    \includegraphics[width=\columnwidth]{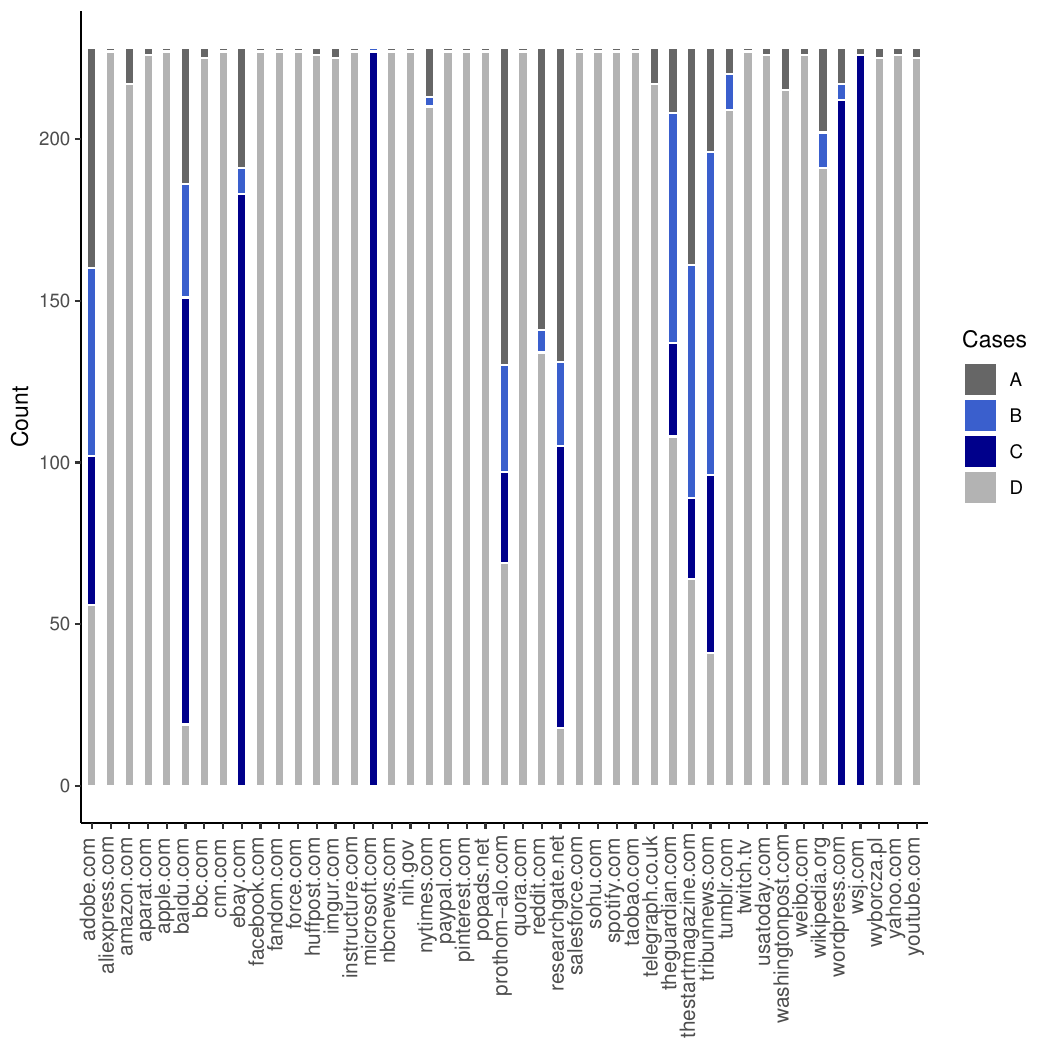}
    \caption{A, B, C, D cases across websites.}
    \label{fig:websites}
\end{figure}

\reffig{fig:websites} shows the frequency of the A, B, C, and D cases on the y-axis against website URL on the x-axis.
The figure shows cases based on the 2019 network performance data across all network providers and regions.
We observe that most of the cases are in the A and D categories, which implies that website performance would improve with lower network latency.
However, there is also a significant percentage of B and C cases. 
Specifically, B and C cases represent 5.69\% and 9.02\% of cases respectively.
For 8 out of 45 (18\%) of the websites B and C cases represent the majority of cases, which means that these websites are limited by insufficient bandwidth in most network/region combinations.
This result shows a contradiction to the accepted wisdom that website performance is limited by insufficient latency in general.
While that is true for most of the measured websites, it is not true for all.
In fact 15 out of 45 (33\%) of these websites are limited by network bandwidth in at least some network/region performance envelopes.

Our recommendation is that developers evaluate the CPI for their websites and compare it against network performance to determine whether the websites are bottlenecked on latency or bandwidth in a given network.
In \refsec{sec:discussion} we discuss techniques they may apply, depending on its limiting factor in a given website, to take advantage of available network performance.

\begin{figure}
    \centering
    \begin{subfigure}{\columnwidth}
        \centering
        \includegraphics[width=0.6\columnwidth]{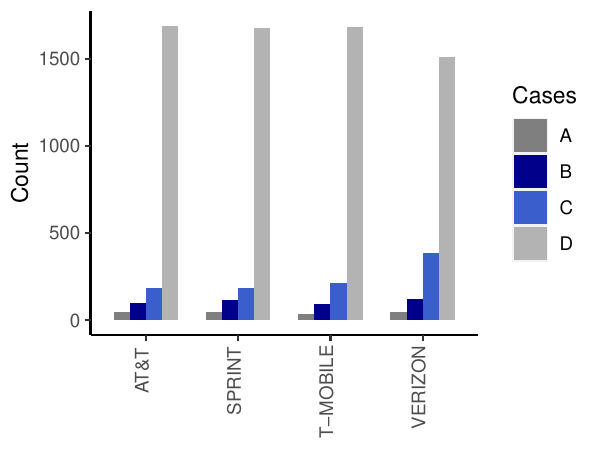}
        \caption{Network data from 2018~\cite{2018OpenSignal}.}
        \label{fig:cpu}
    \end{subfigure}%
    
    \begin{subfigure}{\columnwidth}
        \centering
        \includegraphics[width=0.6\columnwidth]{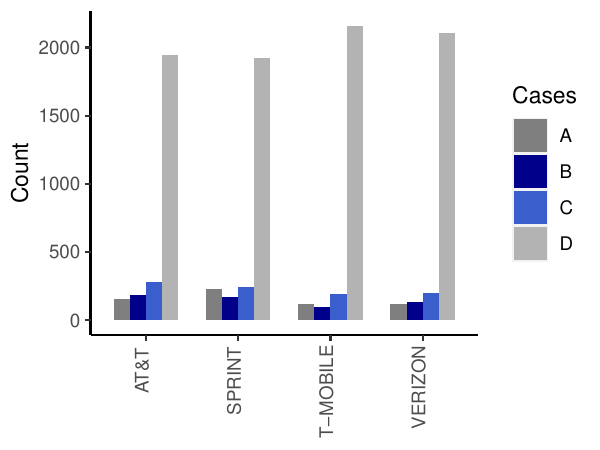}
        \caption{Network data from 2019~\cite{2019OpenSignal}.}
        \label{fig:memory}
    \end{subfigure}
    
    \begin{subfigure}{\columnwidth}
        \centering
        \includegraphics[width=0.6\columnwidth]{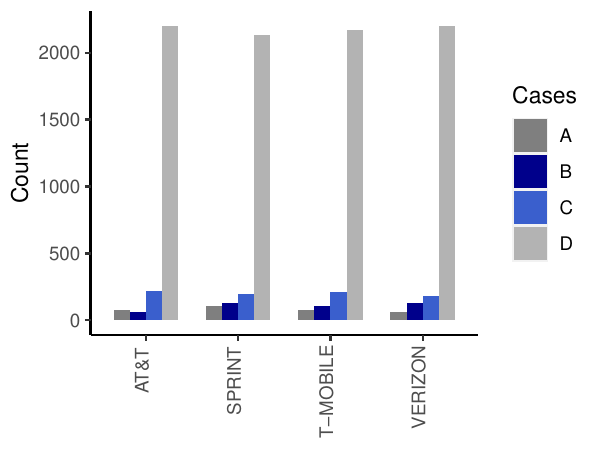}
        \caption{Network data from 2020~\cite{2020OpenSignal}.}
        \label{fig:network}
    \end{subfigure}
    
    \caption{A, B, C, and D cases across network providers.}
    \label{fig:providers}
\end{figure}

\begin{figure*}[h!]
    \centering
    \begin{subfigure}{\textwidth}
        \centering
        \includegraphics[width=1\textwidth]{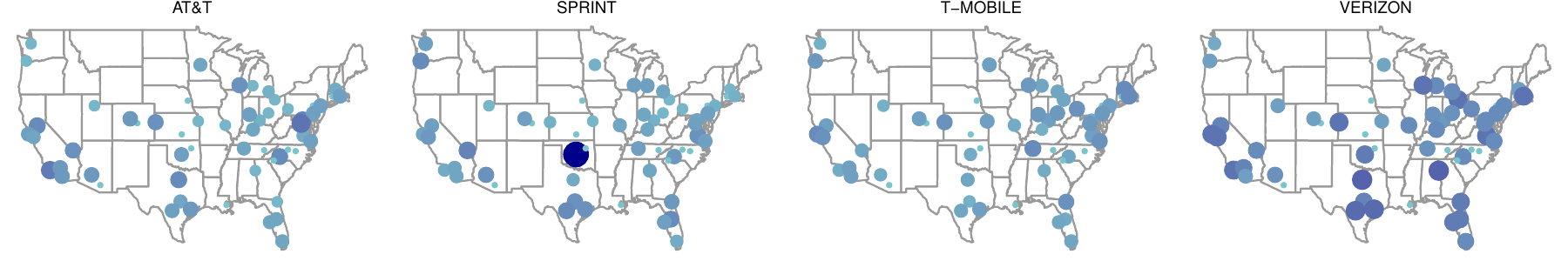}
        \caption{Network data from 2018~\cite{2018OpenSignal}.}
        \label{fig:geography:2018}
    \end{subfigure}%
    
    \begin{subfigure}{\textwidth}
        \centering
        \includegraphics[width=1\textwidth]{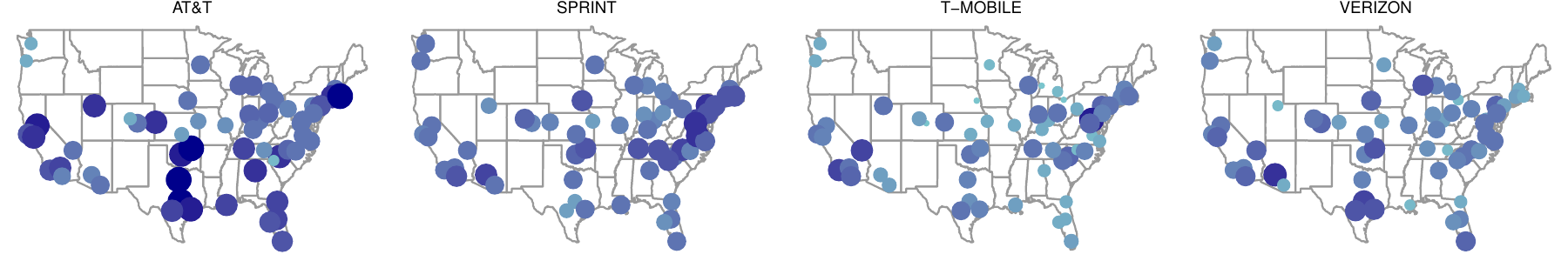}
        \caption{Network data from 2019~\cite{2019OpenSignal}.}
        \label{fig:geography:2019}
    \end{subfigure}
    
    \begin{subfigure}{\textwidth}
        \centering
        \includegraphics[width=1\textwidth]{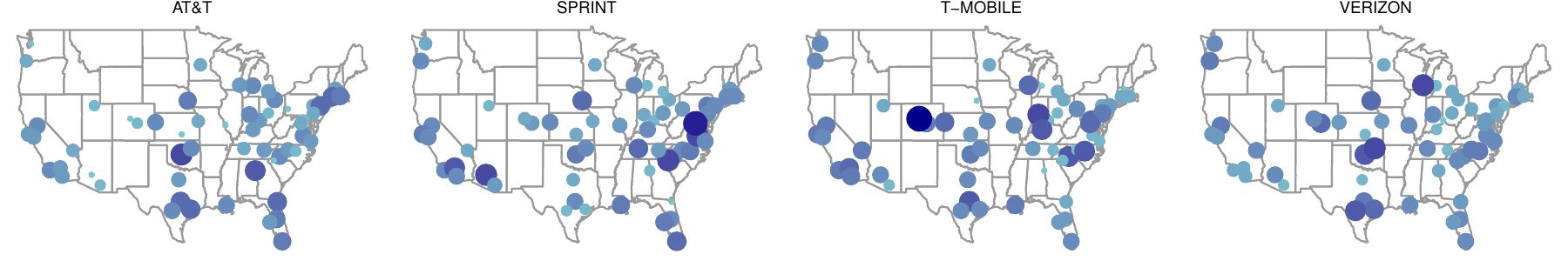}
        \caption{Network data from 2020~\cite{2020OpenSignal}.}
        \label{fig:geography:2020}
    \end{subfigure}
    
    \caption{Ratio of $\frac{B+C}{A+D}$ cases in US cities.}
    \label{fig:geography}
\end{figure*}

\subsubsection{Are certain providers more bandwidth limited?}

We also wanted to understand whether the frequency of B and C cases varies by network provider, or over time.
\reffig{fig:providers} shows the number of A, B, C, and D cases on the y-axis and network providers on the x-axis. 
The three panels show the number of occurrences for the different years of network performance data.

In general, we observe that for all providers most of the cases are in the D and A categories.
At the same time all providers show a number of B and C cases.
The frequency of B and C cases does not vary substantially across the years and in fact has decreased over the years as network providers upgrade their infrastructure.
Still in the 2020 network data, B and C cases represent 12.12\% of cases.
The number of B and C cases does not differ substantially across providers, with the difference between 371 on T-Mobile with the most B and C cases in 2020 and 314 on Verizon with the fewest being less than 16\%.

\subsubsection{Do bandwidth limitations vary geographically?}

We also wanted to understand whether website performance is limited by bandwidth more often in certain areas of the country.
In \reffig{fig:geography} each point shows the ratio of the occurrence of B and C cases to A and D cases for different US cities.
Each point shows the fraction of $\frac{B+C}{A+D}$ cases with darker/larger points showing higher values, in other words the prevalence of bandwidth-limited websites.
The rows of maps shows data for different years of network performance, while the columns show data for different network providers.

In general we observe that instances of websites being bandwidth limited (darker points showing higher $\frac{B+C}{A+D}$ ratios) as local phenomena that vary by year and provider.
In 2018, websites were bandwidth limited primarily in Oklahoma in the Sprint network and in different regions in the Verizon network.
In 2019, websites were bandwidth limited to a larger degree across all providers and regions.
In 2020, as network performance improved, bandwidth bottlenecks were more prevalent in North East and the South West regions of the country for all providers.

While providers in strive to improve network performance both in terms of latency and bandwidth, they may not be aware how different websites perform in their networks regionally.
By understanding the prevalence of the different types of cases in different regional network deployments, providers may adjust network resource allocation to better meet website needs.

\subsubsection{Do small changes to bandwidth and latency improve PSI?}

\begin{figure}
    \centering
    \includegraphics[width=0.7\columnwidth]{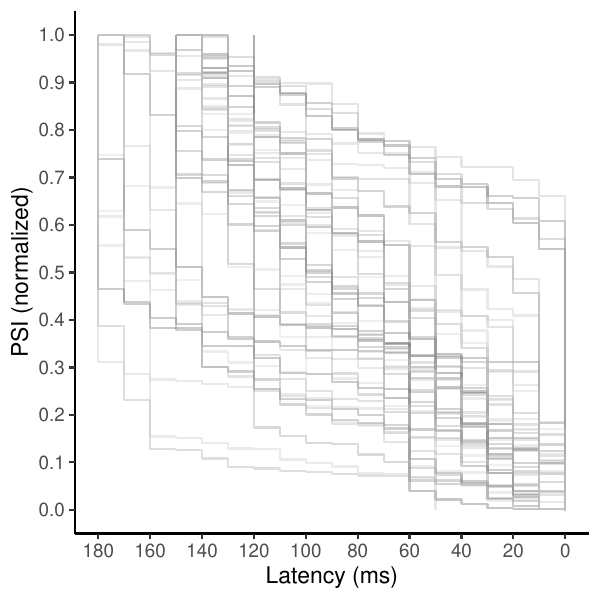}
    \caption{PSI versus latency.}
    \label{fig:psi_vs_latency}
\end{figure}

\begin{figure}
    \centering
    \includegraphics[width=0.7\columnwidth]{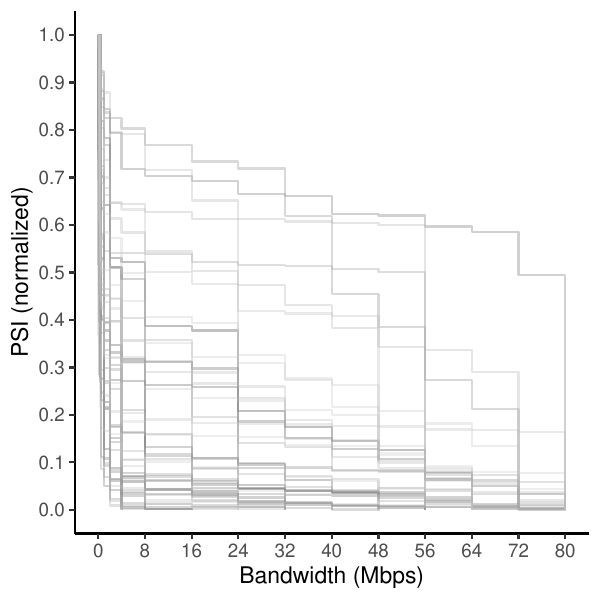}
    \caption{PSI versus bandwidth.}
    \label{fig:psi_vs_bandwidth}
\end{figure}

We also wanted to understand how small changes to network latency and bandwidth alone improve website performance.
In \reffig{fig:psi_vs_latency} we show the effect of network latency on the x-axis on PSI on the y-axis.
In \reffig{fig:psi_vs_bandwidth} we show the effect of network bandwidth on the x-axis on PSI on the y-axis.
The lines in the graphs show page CPIs.

In general we observe that as network performance improves so does the PSI.
Somewhat contrary to results observed by Belshe and Grigorik~\cite{Belshe2010More, Grigorik2012Latency} we observe that both improvements in latency and improvements in bandwidth have a steady effect on page PSI.
In other words we do not observe the diminishing improvements to page performance of additional bandwidth for all pages.
This result is interesting in that as modern pages become more complex, there is no clear path to improving page performance by simply decreasing network latency. 
Instead a careful analysis of page CPI and network performance gives more insight.

\subsubsection{What makes websites limited by bandwidth?}

We also wanted to understand what aspects of a page make it more likely to fall into the B and C cases.
To do so, we use our measurement data to predict the ratio of B and C cases to A and D cases ($\frac{B+C}{A+D}$) using page characteristics.
To construct the data set for our model we compute the $\frac{B+C}{A+D}$ ratio for each page, provider network, and year.
Next, we compute page characteristics as the number of script~(\ttt{.js}), Cascading Style Sheets~(CSS)~(\ttt{.css}), image~(\ttt{.png, .jpg, .gif, .svg}), and text~(\ttt{.html, .json}) files as well as their total size in KB, from HTTP Archive~(HAR) files for each page. 
Based on these data we train a linear regression model~({\footnotesize\texttt{sklearn.linear\_model.LinearRegression}}) using an 80/20 training/testing split in a 100 trials.

\begin{table}[h]
\centering
 \begin{tabular}{l l l l l} 
 \hline
 Factor & Coefficient Value \\ 
 \hline
 \hline
 CSS count &                     0.261606 \\ 
 image count &                   0.041035 \\
 script count &                  0.037109 \\
 CSS KB &                        0.025595 \\
 text count &                    0.019713 \\
 text KB &                       0.002346 \\
 script KB   &                   0.000432 \\
 image KB &                      0.000123\\ 
 \hline
\end{tabular}
\vspace{10pt}
\caption{Page factors in linear regression of the $\frac{B+C}{A+D}$ ratio.}
\label{tbl:factors}
\end{table}

On average we are able to predict the $\frac{B+C}{A+D}$ ratio withing about a factor of 2.4 on average.
While the behavior of pages is complex and depends on their structure, this prediction nevertheless gives us some insight into the influence of the page characteristics that results in particular $\frac{B+C}{A+D}$ ratios.
\reftbl{tbl:factors} shows the model factors (page characteristics) and their sorted coefficient values.
We observe that the number of CSS, image, and script objects have the greatest effect on the $\frac{B+C}{A+D}$ ratio.
When comparing page $\frac{B+C}{A+D}$ ratio to the number of CSS, image, and script objects on a page we observe that page/network combinations with high $\frac{B+C}{A+D}$ ratios tend to have more of these objects. 
However, there are also pages which fall into D cases in all networks with similar number of objects.
Thus, we conclude that a more in-depth analysis of page structure, which results in higher $\frac{B+C}{A+D}$ ratios, should be a part of future work.
In the meantime, we reiterate, that developers should measure the performance of their pages in different network conditions directly to understand if there are network scenarios where their pages are limited by bandwidth.

\section{Discussion}
\label{sec:discussion}

Our results show a mismatch between website needs in terms of network performance, expressed through the CPI, and the network performance offered by network operators.
Within each network performance envelope a website is limited by either insufficient bandwidth or latency.
If a website is bandwidth-limited within a network performance envelope, it could improve user experience by reshaping its CPI through page optimization.
Alternatively, if a website is latency-limited it could be redesigned to change its CPI and take advantage of the extra bandwidth available in a given network.
Below we offer a discussion of techniques website developers could apply to their sites to change their CPI to better fit available network performance.
While the optimization approaches discussed below are already used by website developers and CDNs, we think that grouping them by impact is of value to design strategies for reshaping page CPIs.

If a website is limited by high latency, website developers could leverage different CDNs to serve content from replica servers closer to users in a given network~\cite{Nygren2010Akamai}.
If possible, they could move origin logic onto the CDN for even faster delivery.
Developers could also use Transport Layer Security~1.3 for faster connection setup times~\cite{Rescorla2018Transport}.
Finally, they could deliver their content using HTTP/3 over the QUIC protocol to tackle lossy networks and retransmission~\cite{Bishop2020Hypertext}.

If a website is limited by low bandwidth, developers could use HTTP link headers to preload CSS and script assets critical for rendering embedded objects without having to wait for the browser to identify these resources and prioritize their HTTP requests~\cite{Osmani2017Preload}.
This approach helps to reduce network idle time as the browser parses page HTML.
Developers could also defer loading of non-critical assets after the onload event, which will improve page interactivity, or defer below-the-fold assets, since they do not contribute to lower the PSI.
Developers may also adapt image sizes to the quality of the network.
For example, Akamai Image Manage and Adaptive Image Compression dynamically adjusts image sizes in response to network quality~\cite{Akamai2020Image, Akamai2020Adaptive}.

\section{Conclusions and Future Work}
\label{sec:conclusions}

We investigated the degree to which latency continues to be the limiting factor to mobile website performance.
Through a novel webpage performance characterization method, Critical Path of Improvement~(CPI), we show the sensitivity of websites to small changes in network bandwidth and latency performance. 
Comparing against measurements of mobile network performance we show instances where websites do suffer from a lack of bandwidth in mobile networks.
Finally, we suggest techniques that web developers and network administrators can use to create better matches between website requirements and network performance by reshaping page CPI.

To be sure, CPI should be used to investigate factors other than bandwidth/latency on website performance. 
Two interesting directions are to investigate the impact of website design features on CPI and of web server configurations, including different network protocols used to serve web content. 
While we consider such investigations a part of future work and hope that CPI will be a useful tool there, we believe they are out of scope for this paper, which focuses on taking a fresh look at the predominance of latency as the limiting factor in website performance.

Another interesting direction for future investigations is to incorporate the effects of jitter and packet loss in CPI characterization.
Jitter and loss could arise due to packet queuing, mobile network scheduler decisions, and route switching.
To keep our discussion focused, we do not explicitly control for these factors, choosing instead to observe them in wild though measurements of page PSI in live networks.
However, CPI could be measured on axes other than latency/bandwidth and opens a path to including jitter and loss in future work.

\section*{Disclosure}

The positions, strategies, or opinions reflected in this article are those of the authors and do not necessarily represent the positions, strategies, or opinions of Akamai.

\small
\bibliographystyle{references/splncs03_unsrt}
\bibliography{ref}

\end{document}